\documentclass[aps,prf,reprint,onecolumn]{revtex4-2}
\usepackage[usenames,dvipsnames]{xcolor}
\usepackage{lipsum}
\usepackage{graphicx}
\usepackage{array}
\usepackage{epstopdf}
\usepackage[justification=justified]{caption}
\usepackage{amsmath}
\usepackage{multirow}
\usepackage{amsmath}
\usepackage{amsfonts}
\usepackage{amssymb}
\usepackage{mathrsfs}
\usepackage{mathtools}
\usepackage{tikz}
\usepackage{pgfplots}
\usepackage{csquotes}

\definecolor{color1}{RGB}{0,113,188}
\definecolor{color2}{RGB}{216,82,24}
\definecolor{color3}{RGB}{236,176,31}

\DeclarePairedDelimiter{\evdel}{\langle}{\rangle}

\begin{document}

\author{Igor A. Maia}
\author{Peter Jordan}
\affiliation{Pprime Institute, CNRS, Universit\'e de Poitiers, ENSMA, Poitiers, France}
\author{Andr\'e V. G. Cavalieri}
\affiliation{Divis\~ao de Engenharia Aeron\'autica, Instituto Tecnol\'ogico de Aeron\'autica, S\~ao Jos\'e dos Campos, Brazil}

\title{Wave cancellation in jets with laminar and turbulent boundary layers: the effect of nonlinearity}

\begin{abstract}


This paper presents a study on wave cancellation in forced jets. Building on recent work on real-time control of forced turbulent jets \cite{Maiaetal2020_control,MaiaetalPRF2021}, we here assess the effect of jet upstream conditions and nonlinearity on wave-cancellation performance. The experiments are performed in jets with laminar and turbulent boundary layers inside the nozzle. An open-loop campaign is first conducted, in which the goal is to analyse the jet response to stochastic forcing with variable bandwidth. The upstream conditions of the jet are found to have a strong influence on the jet response. For narrow forcing bandwidths, both jets present a clear response regime. However, in the initially-laminar jet, as bandwidth is increased, high growth rates and transition to turbulence in the initial region underpin the onset of nonlinear effects in jet response. In the initially-turbulent jet, on the other hand, lower growth rates allow a linear response regime to be maintained for a broader range of forcing parameters. As the wave cancellation strategy is linear, reactive control is found to be more effective in the initially turbulent jet, consistent with the results of the open-loop analysis. 

\end{abstract}

\maketitle

\section{Introduction}
\label{intro}
Manipulating flow characteristics to achieve a desired objective is a pressing challenge in technological applications. Flow control applications include, for example, reducing drag in streamlined and bluff bodies, optimising heat transfer in combustion chambers, reducing structural vibration and reducing aerodynamic noise radiation by turbulent jets. Some of these problems involve unsteady flow dynamics, which makes them all the more challenging. This is the case for the jet-noise problem.

Attempts to reduce jet noise have often relied on passive and open-loop devices. These may involve, for example, nozzle modifications in the form of \textit{tabs} and \textit{chevrons} \citep{ZamanBridges2011} or beveled nozzles \citep{ViswanathanBeveled}. Alternatively, fluidic actuation such as microjets \citep{Henderson2010, Maury2012, KoenigJFM2016} or plasma \citep{Samimyetal2004} have also been studied. For subsonic jets, these solutions can achieve up to 3dB noise reduction in low frequencies \citep{ZamanBridges2011}, but are usually followed by an increase in high frequencies and/or a penalty in thrust (in the case of \textit{tabs} and \textit{chevrons}), limiting their application in commercial aircraft. Furthermore, their development often contains much empiricism and trial-and-error, due to the complexity of the jet-noise problem.

In that sense, knowledge about the dynamics of coherent structures, also referred to as wavepackets, in jets and their now demonstrated importance for the sound field \citep{JordanColoniusReview,CavalieriatalAMR2019} can be used to shed light on the mechanisms of sound reduction of open-loop devices and thereby aid control design. Wavepacket modelling through linear stability theory has been used, for example, to understand passive control mechanisms on a flow over a backward-facing step \citep{OrmondeEXPIF2018}, to clarify the effects of chevrons in the mean flow stability and sound radiation of subsonic jets \citep{SinhaJFM2016}, and to understand noise reduction achieved by fluidic actuation from a rotating plug  \citep{KoenigJFM2016}. Linear theory can also be used to understand the sensitivity of wavepackets to external forcing, as done by \citet{TissotJFM2017}, and provide guidance for actuator placement.

Another approach is closed-loop/reactive flow control, wherein flow measurements are used in real-time to produce an unsteady actuation signal via a control law. Due to the nonlinearity and high-dimensionality of fluid systems, the design of control strategies often relies on reduced-order models. These can be obtained, for example, through the linearisation of the equations of motion, or by Galerkin projection approaches based on non-linear equations. Linear stability theory then appears as a candidate to provide control-law designs \citep{Bagherietal2009}. Alternatively, control laws can be obtained by means of transfer functions obtained through empirical system-identification techniques \citep{HerveetalJFM_ARMAX,SasakiJFM2017}.

In a linear framework, disturbances can be eliminated through a simple superposition of waves in a destructive pattern, in a process referred to as \textit{wave cancellation}. This approach has proved successful to control flows underpinned by linear instability mechanisms such as laminar boundary layers \citep{Thomas&Saric1981, Thomas1983,Laurien&Kleiser1989,Li&Gaster2006, SasakiTCFD2018_2} and transitional mixing layers \citep{SasakiTCFD2018_1}.

In the case of the turbulent jet, wave-cancellation-based control is also appealing insofar as many important traits of jet dynamics are governed by linear mechanisms \citep{JordanColoniusReview,CavalieriatalAMR2019}. This is the subject of the present study, wherein we perform an experiment to control artificially-forced, axisymmetric disturbances (wavepackets) in jets. The external forcing raises wavepacket amplitudes above the level of background turbulence, facilitating their identification, as shown by \citet{CrowChampagne} and \citet{Moore1977}, and their control.

The success of this approach relies on the linearity assumption. For harmonically-forced disturbances, \citet{CrowChampagne} and \citet{Moore1977} have shown that it is possible to obtain a linear response regime for a certain range of forcing amplitudes, creating the necessary condition for wave cancellation. Indeed, wave cancellation of harmonic disturbances in jets has been performed both in open-loop (\cite{KopievAcPhy2013}) and in closed-loop (\cite{KopievAIAA2019_1})configurations. Recently, \citet{Maiaetal2020_control, MaiaetalPRF2021} have provided a proof-of-concept that real-time reactive control of stochastic disturbances through wave cancellation is also possible in fully turbulent jets.

Here we build on those studies and analyse the impact of initial conditions on the linearity of the jet response and on the effectiveness of wave cancellation. We first conduct open-loop measurements to find out to what extent the jet response to stochastic forcing can be maintained in a linear regime. This is done in jets with both laminar and turbulent exit boundary layers by imposing stochastic signals of variable bandwidth and amplitude. Then, based on a simplification of the control law proposed by \citet{SasakiTCFD2018_2}, we perform wave-cancellation of the said disturbances in real-time and use the open-loop results to interpret control performance.

The remainder of the paper is organised as follows. In section \S \ref{sec:exp_setup} we present the experimental setup and describe the flow conditions adopted. In \S \ref{sec:stability} we show results of a locally parallel linear stability analysis performed on the initially-laminar and initially-turbulent jets. This is followed by a description of the control law design in \S \ref{sec:control_design}. In \S \ref{sec:results} we show results of the open-loop and reactive experimental campaigns and we finish with conclusions in \S \ref{sec:conclusions}.

\section{Experimental setup}
\label{sec:exp_setup}

The experiments were performed in a low-Mach-number facility at the Pprime Institute, in Poitiers, France. The jet is fed by a centrifugal fan equipped with an air filter and exits through a straight nozzle of diameter $D=50$mm. A settling chamber featuring a honeycomb panel and mesh grids is located upstream of the nozzle and is followed by a convergent section with a contraction ratio of 25. The jet velocity was adjusted by a variable-frequency driver actuating on the fan and the room temperature was monitored so as to ensure constant Mach number conditions. The jet Mach ($Ma=U_{j}/c_{\infty}$) and Reynolds ($Re=U_{j}D/\nu$) numbers are $0.05$ and $5 \times 10^4$, respectively, with $U_{j}$ the jet exit velocity, $c_{\infty}$ the ambient speed of sound and $\nu$ the kinematic viscosity of the air.

\begin{figure}[!ht]
\centering
\includegraphics[trim=0cm 0cm 0cm 0cm, clip=true,width=0.4\linewidth]{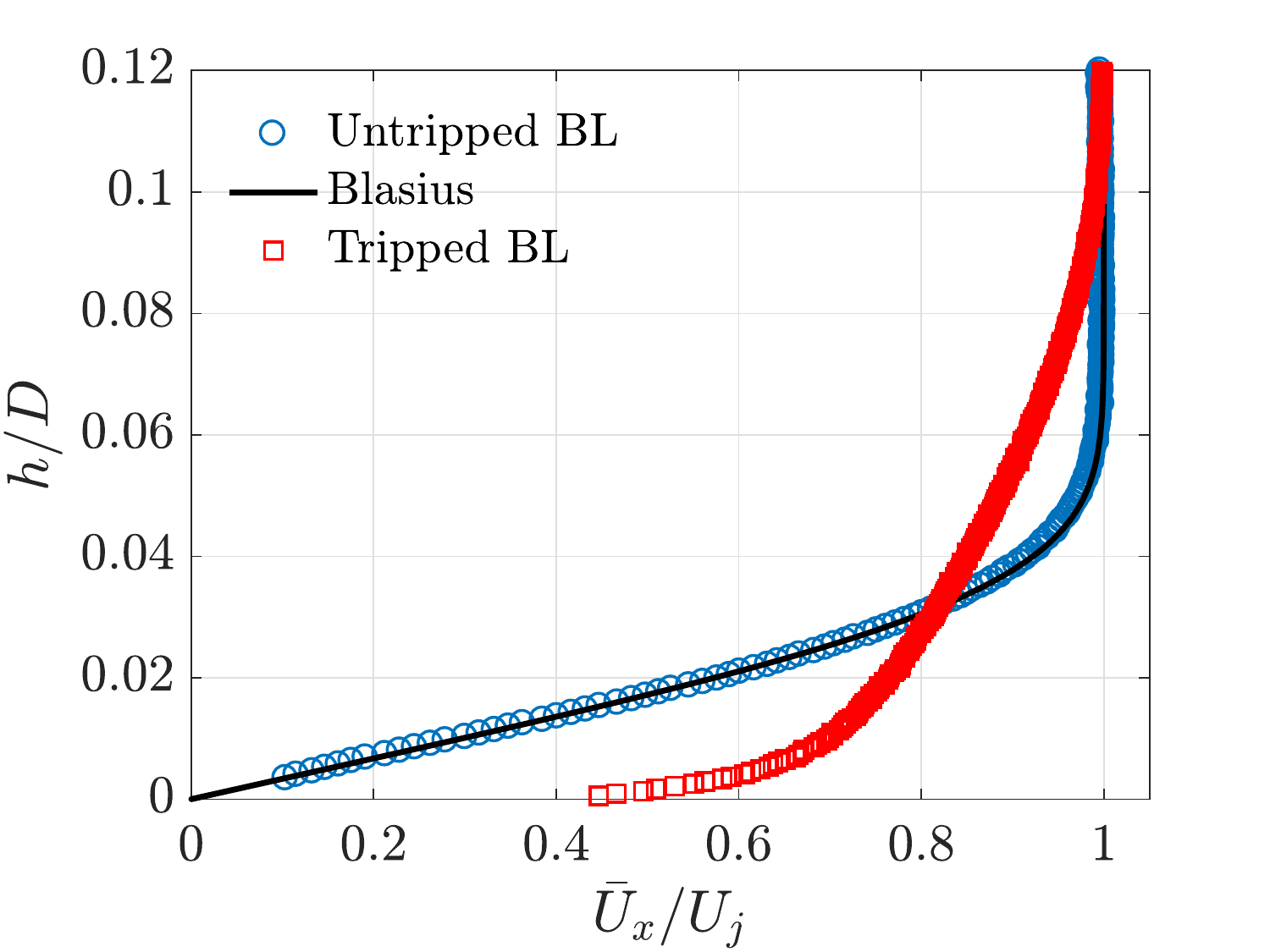}
\caption{Boundary layer profiles at the nozzle exit plane. The untripped boundary layer is compared to the Blasius solution for a laminar boundary layer. $\bar{U}_{x}$ is the mean streamwise velocity and $h$ is the normal distance from the wall.}
\label{bl_comp}
\end{figure}

Two different nozzles were used: one with an untripped boundary
layer, with a smooth interior surface; and another one in which the boundary layer was tripped $2.5D$ upstream of the nozzle exit using carborundum particles. Jets issuing from both nozzles were characterised by hot-wire and Particle Image Velocimetry (PIV) measurements. The hot-wire was a Dantec 55P11 model the PIV system consisted of two Photron APS-RS cameras and a 527 nm 30mJ Continuum TERA PIV laser. The PIV measurements were performed in a plane parallel to the jet axis, and the cameras covered the range $-1.5\leqslant r/D \leqslant 1.5$, $0 \leqslant x/D \leqslant 6$, where $r$ and $x$ are the radial and streamwise coordinate, respectively. The sampling frequency of the PIV system was limited to 1.5kHz, and the measurement time was set to 2700 convective time units, with a convective time unit defined by $D/U_{J}$. PIV computations were carried out using a commercial software which performed a multi-pass iterative PIV algorithm \citep{Scarano2002}. The PIV interrogation area size was set to 64x64 pixels for the first pass, decreased to 16x16 pixels with an overlap of 50 \% between two neighbouring interrogation areas.

Figure \ref{bl_comp} shows boundary layer mean velocity profiles measured with the hot wire. The untripped boundary layer is compared to the Blasius solution for a laminar boundary layer over a flat plate; the agreement is excellent, showing that the untripped nozzle produces an initially laminar jet. In Appendix \ref{sec:turb_bl} we present a comparison between tripped boundary layer profiles and data from the literature for turbulent boundary layers. The characteristics of the tripped boundary layer are consistent with those of a fully turbulent flow, showing that the tripping successfully provokes transition to turbulence. Table \ref{tb_bl} shows the boundary layer thickness, $\delta$, and momentum thickness, $\delta_{2}$ for all both flow conditions.

\begin{table}[!hb]
\centering
\begin{tabular}{c c c c}
\hline
\multicolumn{4}{c}{Untripped BL}\\
\hline
$\delta$ (mm) & $\delta/D$ & $\delta_{2} (mm)$ & $\delta_{2}/D$ \\
2.76 & $5.5\times 10^{-2}$ & 0.37 & $7.5\times 10^{-3}$ \\
 \hline
 \multicolumn{4}{c}{Tripped BL} \\
 \hline
 $\delta$ (mm) & $\delta/D$ & $\delta_{2} (mm)$ & $\delta_{2}/D$ \\
4.94 & $1.0 \times 10^{-1}$ & 0.55 & $1.1\times 10^{-2}$ \\
\hline
\end{tabular}
 \caption{Boundary layer thickness, $\delta$, and momentum thickness, $\delta_{2}$.}
 \label{tb_bl}
\end{table}

Mean and rms velocity profiles measured in the jet plume are reported in Appendix \ref{sec:appendix}. The results show that for the jet with the untripped boundary layer rms levels go from very small values in the vicinity of the nozzle to a peak around $x/D=1$ and a decrease further downstream, indicating that transition to turbulence occurs in the first jet diameter (this is also clear from the vortex roll-up seen in the flow-visualisation images and videos made with the PIV system and provided as supplementary material). The jet with the tripped boundary layer, on the other hand, presents rms values consistent with a turbulent state for all positions measured.

\begin{figure}[!ht] 
\centering
\includegraphics[trim=5cm 5cm 5cm 7cm, clip=true,width=0.6\linewidth]{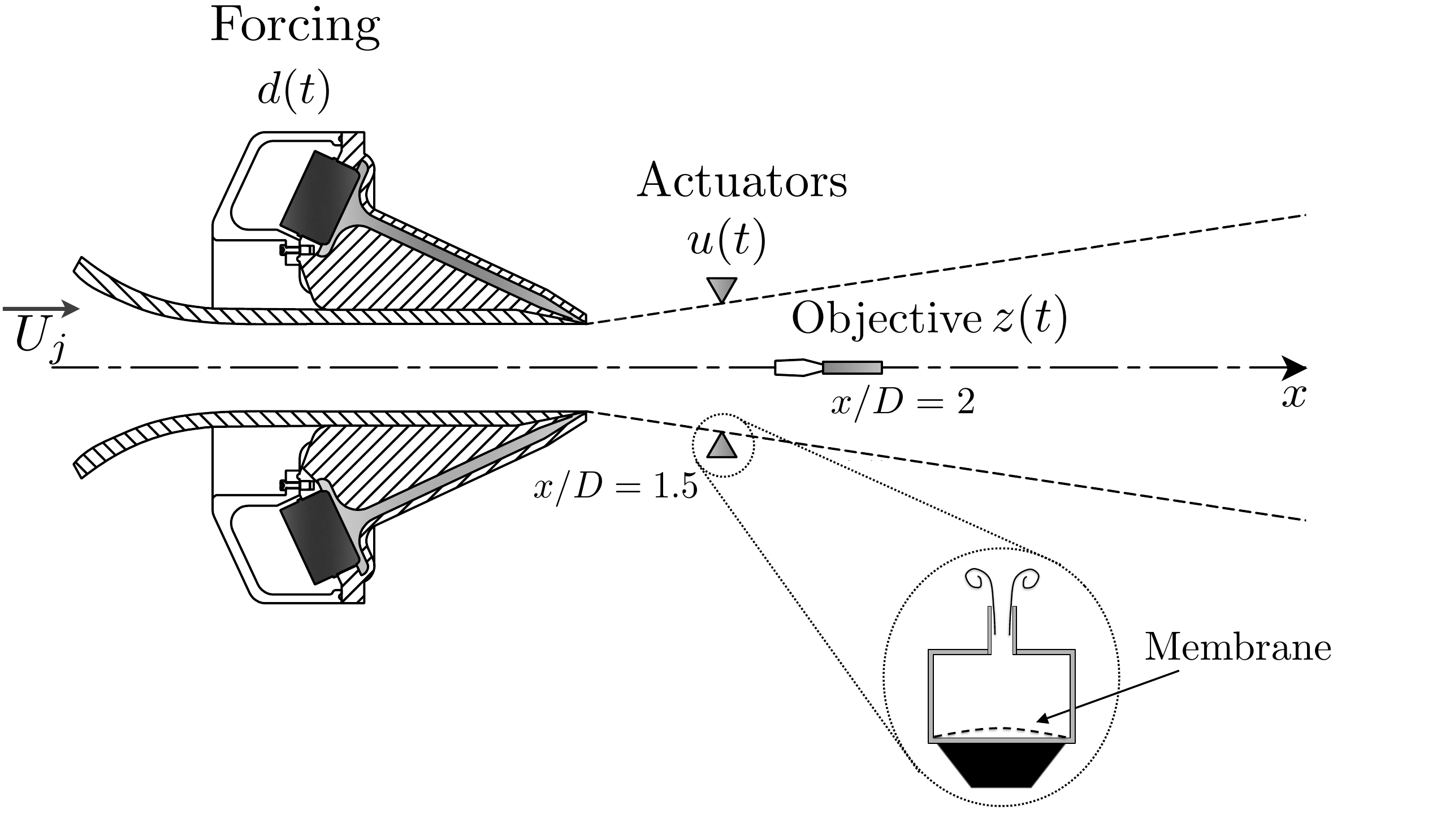}
\caption{Schematic of wave cancellation experiment in forced jets. Forcing and actuation consist of synthetic jets generated by loudspeakers; the objective consists of streamwise velocity measurements performed by a hot wire at the jet centerline.}
\label{setup}
\end{figure}

The wave-cancellation setup involves forcing and actuation systems and an objective, which we seek to minimise. Figure \ref{setup} shows a schematic of the experiment. The forcing, $d$, is produced by a system of eight loudspeakers (model AURA NSW 2-236-8AT) equally distributed in the azimuthal direction and mounted in a conical structure that fits the nozzle. The loudspeakers are synchronised and produce synthetic jets that exit through a $0.01$D annular gap at the nozzle lip and produced axisymmetric disturbances. Forcing amplitude is varied by changing the voltage applied to the loudspeakers. Actuation, $u$, also consisted of synthetic jets generated by synchronised loudspeakers: six speakers (model AURA 1-205-8 A) were used to drive synthetic jets on a ring array placed at $1.5D$ from the jet exit at a radial position immediately outside of the shear layer. The speakers were placed inside 3D printed boxes with internal dimensions designed to accommodate their membranes. The synthetic jets exit through a $1$mm aperture pointed towards the center of the main jet. The objective, $z$, consists of streamwise velocity fluctuations measured by a hot wire at the jet centerline $2D$ downstream of the nozzle exit.

\section{Local linear stability analysis}
\label{sec:stability}

Linear stability theory has been widely used to model the evolution of wavepackets in jets. Stability models based on a linearisation about the mean flow and a locally parallel flow assumption successfully describe many important aspect of jet dynamics\citep{JordanColoniusReview,CavalieriatalAMR2019}. Here we adopt this approach and perform a spatial linear stability analysis, aiming at understanding the differences in the growth rate of disturbances in the initially-laminar and turbulent jets.

In local analysis, the base flow is considered to be homogeneous in the streamwise and azimuthal directions and in time, which leads to disturbances of the form,

\begin{equation}
q'(x,r,\theta,t)=\hat{q}(r)e^{i(\alpha x-\omega t)}e^{im\theta},
\label{disturb}
\end{equation} 
where $q=\left[u_{x},u_{r},u_{\theta},\rho,T\right]^{T}$ is a vector containing, respectively, the three components of velocity, density and temperature, $\alpha$ and $m$ are the wavenumbers in the streamwise and azimuthal directions, respectively, and $\omega$ is the frequency. We also consider the Reynolds decomposition $q(x,r,\theta,t)=\bar{q}(x,r)+q'(x,r,\theta,t)$, where $\bar{q}$ is the mean flow and $q$ is the fluctuation. Since the jets under study are forced by axisymmetric disturbances ($m=0$), we only focus on the evolution of those disturbances. Equation \ref{disturb} and the Reynolds decomposition are introduced into the compressible Navier-Stokes equations written in cylindrical coordinates and linearised about the mean flow. Variations of the mean flow in the streamwise direction are neglected, following the locally parallel hypothesis.

The linearised Navier-Stokes equations can then be recast as a generalised engenvalue problem

\begin{equation}
\textbf{L}\hat{q}=\alpha\textbf{F}\hat{q},
\label{gen_eigenvalue}
\end{equation}
where $\alpha$ is the eigenvalue and $\hat{q}$ the corresponding eigenvector. Viscous terms containing $\alpha^2$ terms are neglected, due to the high Reynolds number of the flow, following \citet{RodriguezEuropean}. The linearised Navier-Stokes equations for an axisymmetric disturbance ($m=0$) are given in Appendix \ref{sec:linearised_NS}. For a given $Re$ and real $\omega$, the evolution of the disturbances in the jet is governed by the sign of the imaginary part of the wavenumber, $\alpha_{i}$: if $\alpha_{i}<0$, the disturbances will grow exponentially in the positive $x$ direction.

\begin{figure}[!ht] 
\centering
\includegraphics[trim=1cm 7cm 1cm 5cm, clip=true,width=0.8\linewidth]{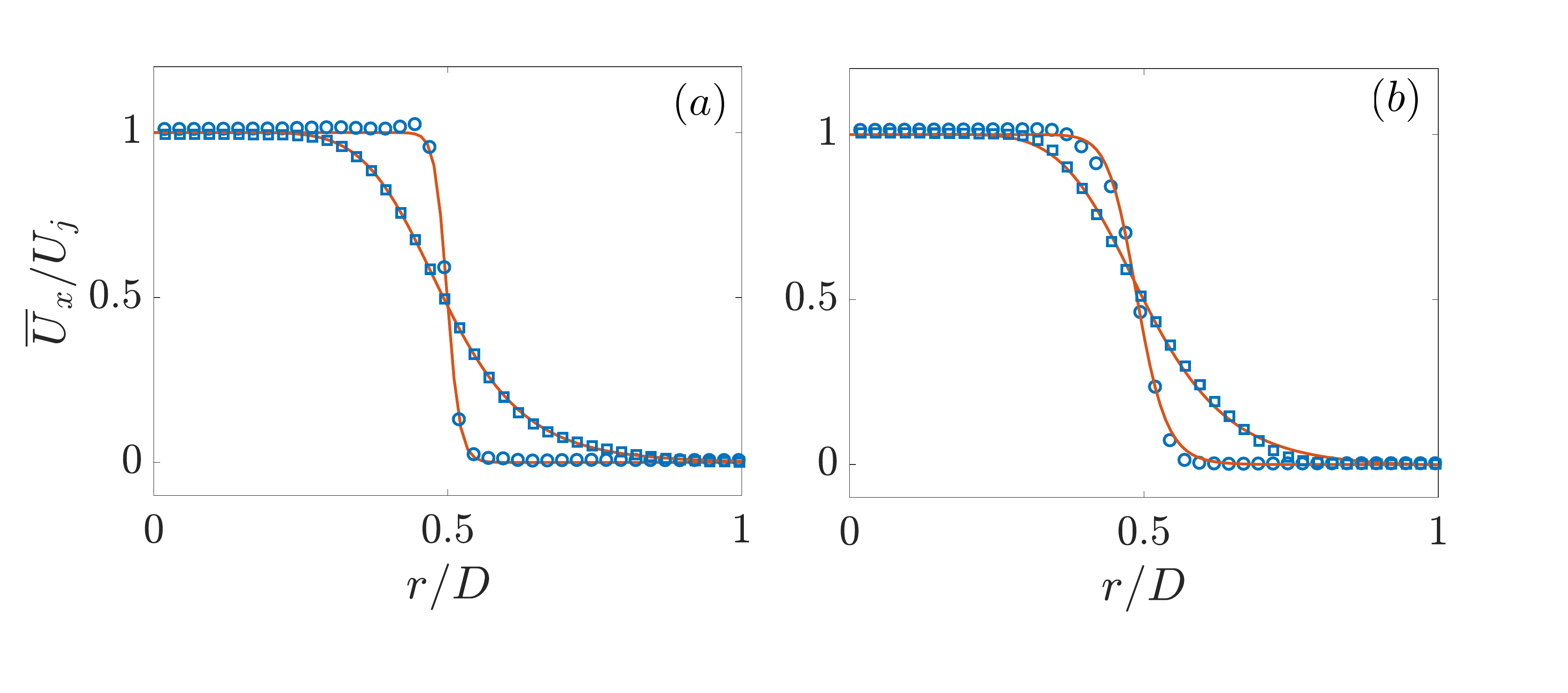}
\caption{Mean flow fit for: (a) Initially-laminar jet; (b) Turbulent jet. The circles correspond to mean experimental profiles measured at $x/D=0.3$ and the squares correspond to measurements at $x/D=2$. The solid lines are the fitted profiles.}
\label{mean_flow_fit}
\end{figure}

Dirichlet boundary conditions are used for a far-field boundary located at $r/D=10^3$ and the radial direction is discretised using Chebyshev collocation points. The domain is extended to the far field by mapping the original domain, $r_{c} \in [-1, 1]$ to $r \in [0,\infty)$ with the function

\begin{equation}
r = \frac{r_{c}+1}{2\sqrt{1-(\frac{r_{c}+1}{2}})^2}
\end{equation}
The system (\ref{gen_eigenvalue}) is then solved for different frequencies and streamwise locations. At each location the mean flow profiles were fitted with the following function:

\begin{equation}
\begin{split}
&U_{fit}/U_{j}= 0.5\biggl\{1+\left[1+c\left(1-\tanh(r)^2\right)\right]\biggl[1+  d\left(\frac{1}{\cosh(r)}\right)^2\biggr]\tanh\left[b\left(\left(\frac{0.5+a}{r}\right)-\left(\frac{r}{0.5+a}\right)\right)\right]\biggr\},
\end{split}
\end{equation}
where $a$, $b$, $c$ and $d$ are constants that have been found through a non-linear least square algorithm. Figure \ref{mean_flow_fit} shows experimental and fitted mean velocity profiles at $x/D=0.3$ and $x/D=2$.

\begin{figure*}[!ht] 
\centering
\includegraphics[trim=0cm 0cm 0cm 0cm, clip=true,width=0.9\linewidth]{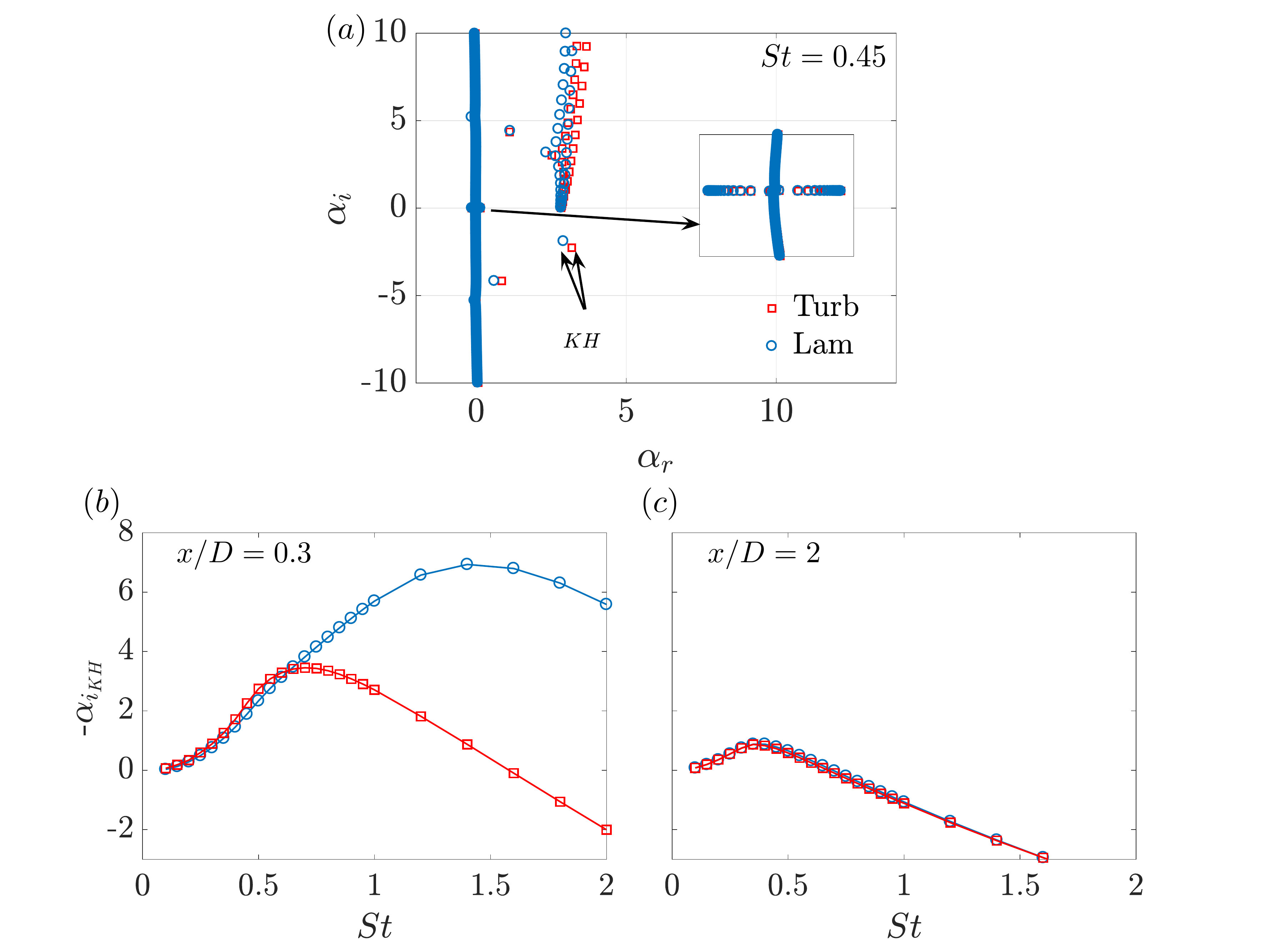}
\caption{Local spatial stability analysis of axisymmetric ($m=0$) disturbances. (a) Typical eigenvalue spectra computed at $St=0.45$ with a mean flow measured at $x/D=0.3$; (b) Growth rate of the Kelvin-Hemlholtz (KH) mode as a function of Strouhal number at $x/D=0.3$; (c) Growth rate of the Kelvin-Helmholtz (KH) mode as a function of Strouhal number at $x/D=2$. Growth rates in (b) and (c) are shown with inverse sign, so that positive values mean exponential growth along positive $x$.}
\label{stability_plots}
\end{figure*}

Figure \ref{stability_plots} shows results of the local spatial stability analysis carried out using the mean flow profiles shown in Figure \ref{mean_flow_fit}. In Figure \ref{stability_plots} (a), typical eigenvalue spectra (computed at $St=0.45$ using the mean flow profile measured at $x/D=0.3$) are shown. Apart from the Kelvin-Helmholtz (KH) modes, six branches can be seen: two branches lying on the real axis, consisting of acoustic modes propagating downstream and upstream; two branches of evanescent modes, one with positive group velocity and another with negative group velocity (both acoustic and evanescent modes are shown in the the zoom of the of Figure \ref{stability_plots} (a)); and two branches of stable modes propagating downstream, one with radial support in the shear-layer and another one with support in the jet core. The pairs of discrete modes lying closer to the acoustic branch (one propagating downstream and the other upstream) are also evanescent, and correspond to acoustic duct-like modes \citep{MartiniJFM2019}.

Figures \ref{stability_plots} (b) and (c) show the growth rates of the KH mode computed at $x/D=0.3$ and $x/D=2$ as a function of Strouhal number. The growth rates are presented with inverse sign, so that positive values mean exponential growth downstream. At $x/D=0.3$, the KH modes in the initially-laminar and initially-turbulent jets were found to have similar growth rate, $\alpha_{i}$ at $St\leqslant 0.65$. However, as the Strouhal number is further increased, the KH mode in the initially-laminar jet displays a significantly higher growth rate. Further downstream, at $x/D=2$ (the objective position in the wave-cancellation experiment), growth rates for the two jets are almost identical. As can be seen, at this streamwise position the jets have become neutrally stable for a range of Strouhal numbers.

It will be seen in section \S \ref{sec:results} that the substantial difference in the growth rates of the jets observed at $St>0.65$ in the initial region has a significant impact on the onset of nonlinear effects when the jet subject to external forcing, thus affecting wave-cancellation performance. 

\section{Wave-cancellation strategy}\label{sec:control_design}

The wave-cancellation strategy is a simplification of that of the inverse feedforward scheme laid out by \citet{SasakiTCFD2018_1} and \citet{SasakiTCFD2018_2}, wherein the objective is expressed as a linear combination of a limited number of flow measurements plus the actuation signal. Here we substitute flow measurements by the forcing signal, $d$. In the frequency domain, the objective is then expressed as:

\begin{equation}
Z(\omega)=D(\omega)H_{dz}+ U(\omega)H_{uz},
\label{output_ctrl_d_freq}
\end{equation}
where $Z$, $D$ and $U$ are the frequency-domain counterparts of $z$, $d$ and $u$. $H_{dz}$ and $H_{uz}$ are the disturbance/objective and actuation/objective transfer functions, defined, respectively as:

\begin{equation}
H_{dz}=S_{dz}/S_{dd},
\label{hdz}
\end{equation}

\begin{equation}
H_{uz}=S_{uz}/S_{uu}.
\label{huz}
\end{equation}
$S_{ij}$ denotes the cross spectral density (CSD) between $i$ and $j$, and $S_{ii}$ the power spectral density (PSD) of $i$.

The actuation signal is expressed as:

\begin{equation}
U(\omega)=K(\omega)D(\omega),
\label{U_ctrl_d}
\end{equation}
where $K(\omega)$ is the control kernel. The expression for the control kernel, which involves the minimisation of an objective functional defined in the frequency domain \citep{SasakiTCFD2018_2}, is given by:

\begin{equation}
K(\omega)=-\frac{H_{dz}}{H_{uz}}.
\label{K_ctrl_d}
\end{equation}
The time-domain actuation signal is given by:

\begin{equation}
u(t)=\int_{0}^{\infty}k(\tau)d(t-\tau)\mathrm{d}\tau,
\label{u_ctrl_d}
\end{equation}
where $k$ is the inverse Fourier transform of $K(\omega)$. In order to avoid noise in regions of the spectrum where $H_{uz}$ has low amplitudes, which might lead to uncontrollable frequencies, kernels are filtered in the frequency bands of forcing and actuation, which will be defined shortly. 

In this approach, $d$ acts at the same time as an external disturbance and an input for the controller. This does not constitute \enquote{classic} closed-loop/reactive control to the extent that no measurements are taken to feed the control law; however it allows us to assess the possibility of performing wave cancellation and understand its limitations. Such a setup, with inputs taken upstream of actuation corresponds to a feedforward configuration, suitable for amplifier flows, such as jets \citep{Sipp&SchmidAMR2016, Schmid&SippPRF2016}. An actual reactive control case, with auxiliary sensors detecting the upstream wavepacket, was considered in \citet{Maiaetal2020_control,MaiaetalPRF2021}, showing significant attenuation for the turbulent jet. This is more relevant for practical applications, but the inclusion of sensors complicates the analysis of linearity that is pursued here. Hence, we focus on the simpler case where the introduced stochastic disturbance is available, in a 	best-case scenario for linear wave cancellation. 

Wave cancellation was performed for the jet forced with stochastic signals filtered in four bandwidths: $0.3 \leqslant St\leqslant 0.45$,  $0.3 \leqslant St \leqslant 0.65$, $0.3 \leqslant St \leqslant 0.85$ and $0.3 \leqslant St \leqslant 1$, where $St$ is the Strouhal number, given by $St=fD/U_{j}$, with $f$ the frequency. The transfer functions were computed in an open-loop step, by measuring the response of the jet to forcing and actuation separately with white-noise signals filtered in the mentioned bandwidths.

A Labview software featuring a real-time module was used to perform the experiments. The software carries out the convolution given by equation \ref{u_ctrl_d} in discrete form at a rate of 5kHz. The generation and acquisition of the signals were made by a National Instruments PXIe-1071 card. We set an acquisition frequency of 30kHz and a measurement time of 30s, which is far superior to the characteristic time scales of the flow, thus assuring the convergence of the single- and two-point statistics necessary for transfer-function computation.

\section{Results}\label{sec:results}
\subsection{Identifying the linear regime}\label{sec:seeking}

The control law is underpinned by two steps: estimation and actuation. In the estimation step, the phases and amplitudes of the forced wavepackets must be predicted at the objective position, via the transfer function $H_{dz}$ of eq. (\ref{hdz}). In the actuation step, the incoming wavepackets are cancelled by wavepackets excited by the actuators with the correct phase and amplitude, assuming that their effect on the objective can be modelled with the $H_{uz}$ transfer function of eq. (\ref{huz}). In both of those steps, the underlying assumption is linearity. Throughout the remainder of the paper, we use two-point coherence, given, for a pair of signals $i$ and $j$ by,

\begin{equation}
\gamma_{ij}^2(\omega)=\frac{|\evdel{S_{ij}(\omega)}|^2}{|\evdel{S_{ii}(\omega)}||\evdel{S_{jj}(\omega)}|},
\end{equation}
as a measure of linearity. The coherence function assumes values between 0 and 1. If $i$ and $j$ have a perfect linear relationship, coherence is equal to 1. If, on the other hand, there is a desynchronisation between the two signals, typical of turbulent flows and nonlinear systems, coherence should decay. 

The two important parameters underpinning the accuracy of the estimation and actuation steps, and therefore control performance, are the disturbance/objective coherence, $\gamma_{dz}$, and the actuation/objective coherence, $\gamma_{uz}$.

It is therefore important, prior to the real-time experiment, to identify a linear response regime associated with forcing and actuation. In the studies of \citet{CrowChampagne} and \citet{Moore1977} such a linear response regime is found to exist if the turbulent jet is forced harmonically. This has motivated a great number of studies dedicated to understanding the flow dynamics and sound field associated with such harmonic disturbances \citep{Kibens1979, Morrison&McLaughlinJSV1979, Favre-Marinet&Binder1979, Jubelin1980, ZamanHussainJFM1981, HoHuerre1984, Raghu, PetersenJFM1991, FreundAIAA2000, SamimyJFM2010, SinhaPhysics2012, CalvoExPIF2014}, and also assessing the possibility of controlling them \citep{SamimyJFM2007, SinhaAIAA2017,KopievAcPhy2013, KopievAIAA2019_1}. Here we build on these works and extend the analysis to stochastically-forced jets.

\begin{figure*}[!ht] 
\centering
\includegraphics[trim=4.5cm 0cm 5cm 0cm, clip=true,width=1\linewidth]{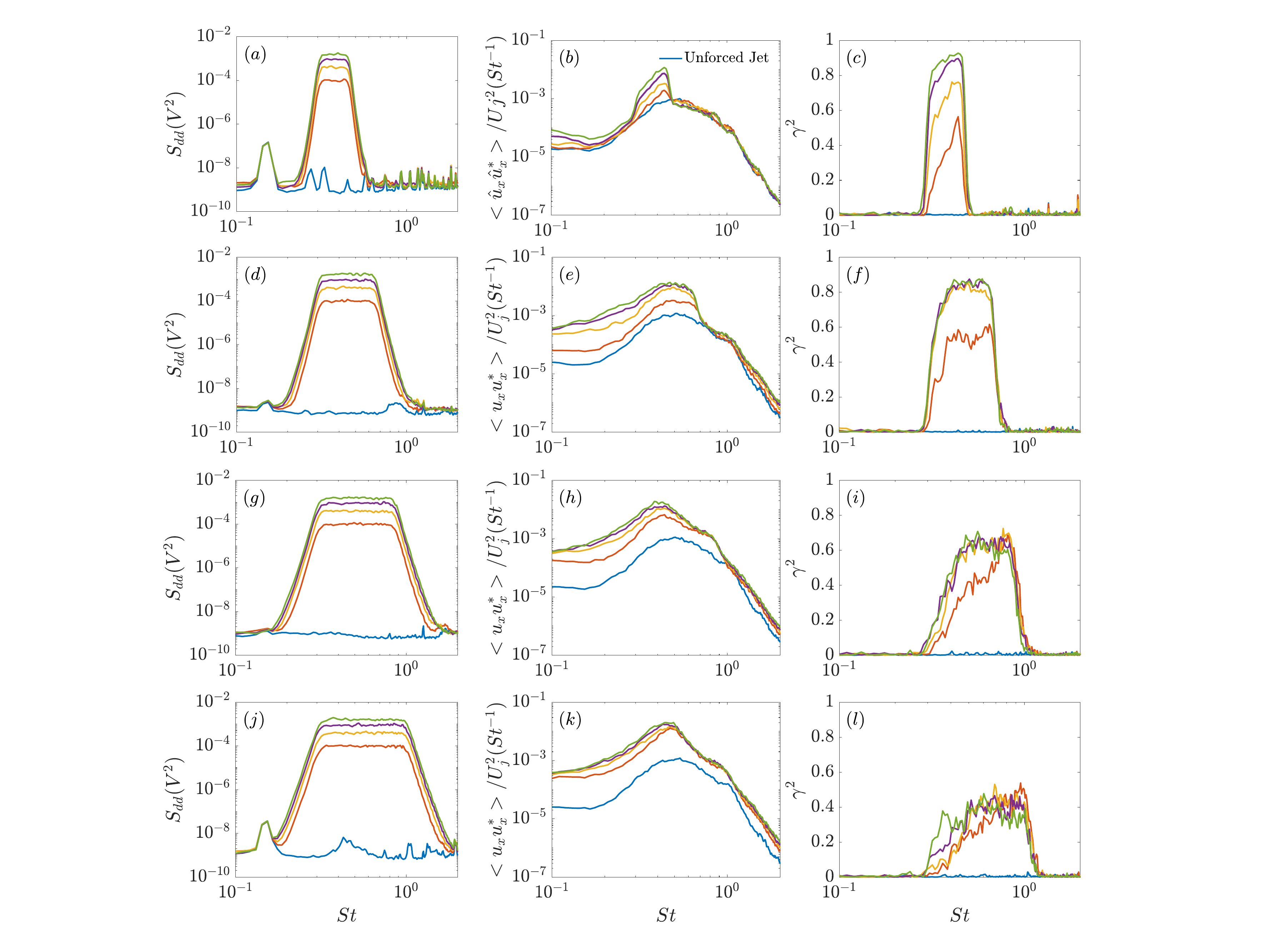}
\caption{Response of the initially-laminar jet to stochastic forcing in increasing bandwidths. Left column: PSD of forcing signals; middle column: PSD of response measured at the objective position; Right column: disturbance/objective coherence. Jets were forced in the following bandwidths: (a)-(c): $0.3\leqslant St\leqslant 0.45$; (d)-(f): $0.3\leqslant St\leqslant 0.65$; (g)-(i): $0.3\leqslant St\leqslant 0.85$; (j)-(l): $0.3\leqslant St\leqslant 1$. The different amplitudes (denoted by the different colors) were kept the same for each forcing bandwidth. Data of the unforced jet is also shown for comparison.}
\label{jet_resp_laminar}
\end{figure*}

In what follows we examine the jet response measured at the objective position, for both the initially-laminar and turbulent cases. Figure \ref{jet_resp_laminar} shows the PSD of forcing and response of the initially-laminar jet, as well as $\gamma_{dz}$, for different amplitudes and bandwidths of forcing. The data from the unforced jet is also shown for comparison. We note that the PSD of velocity fluctuations presents a broadband peak typical of a turbulent jet. This indicates that the jet has already undergone transition at the objective position. The jet displays a clear response over the $St$ range of forcing, with significant amplifications also occurring at lower $St$. For the two narrower bandwidths of forcing, high coherences are attained ($\gamma_{dz}^{2} >0.8$) and there is a systematic increase in the response amplitude with increasing forcing amplitude. However, as the bandwidth is further widened (but keeping the same forcing amplitudes) the response achieves a higher amplitude for a given forcing amplitude; saturation of the response with increasing forcing amplitude occurs more rapidly, and forcing-response coherence is considerably degraded. All of these behaviours can be associated with non-linear effects activated by the increased bandwidth of forcing. 

\begin{figure*}[!ht]
\centering
\includegraphics[trim=4.5cm 0cm 5cm 0cm, clip=true,width=1\linewidth]{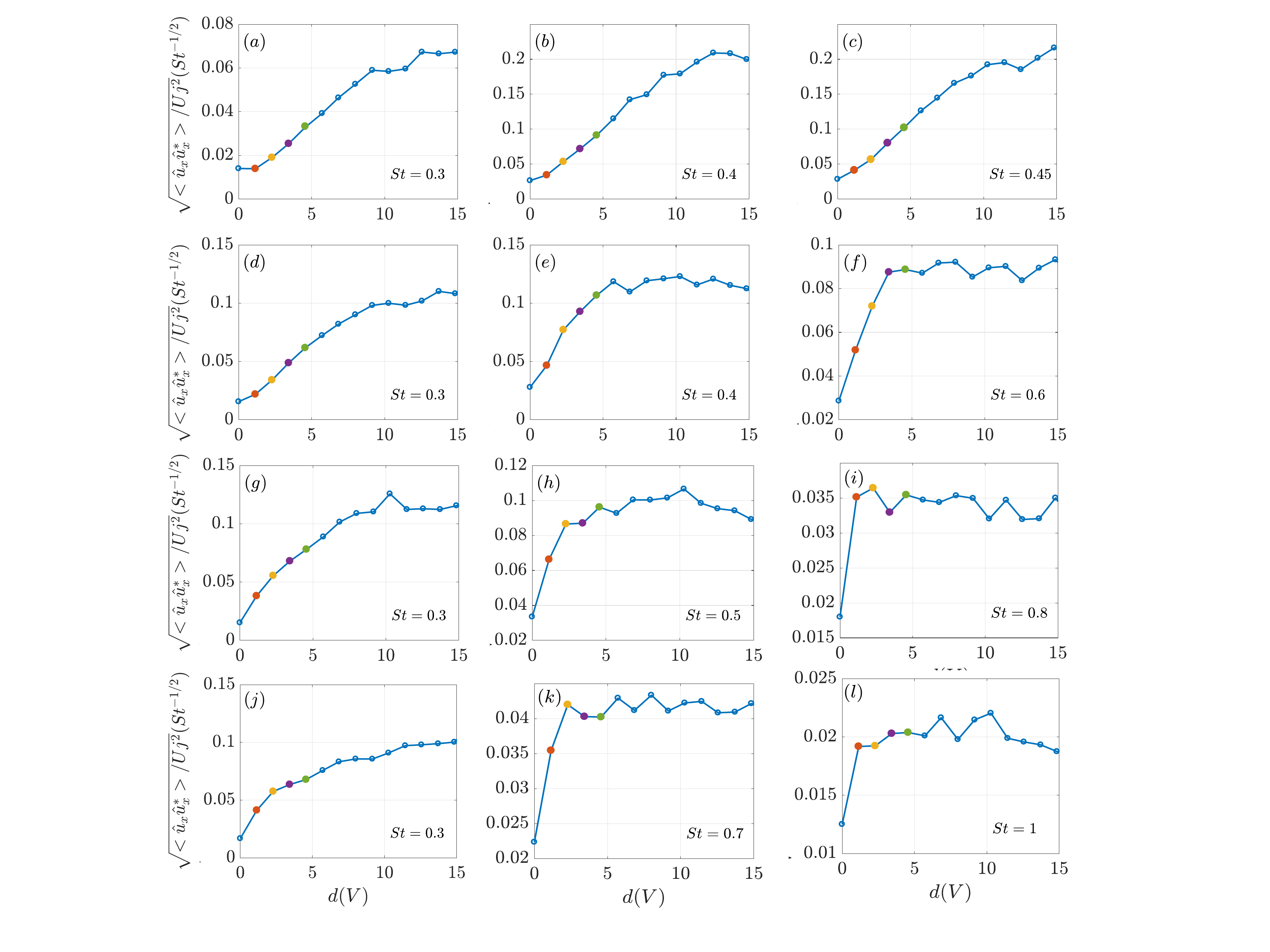}
\caption{Response of the initially-laminar jet as a function of forcing amplitude for selected frequencies spanning the different forcing bandwidths. Plots correspond to forcing in the bands. (a)-(c): $0.3\leqslant St\leqslant 0.45$; (d)-(f): $0.3\leqslant St\leqslant 0.65$; (g)-(i): $0.3\leqslant St\leqslant 0.85$; (j)-(l): $0.3\leqslant St\leqslant 1$. The filled circles correspond to the amplitudes taken from the spectra shown in Figure \ref{jet_resp_laminar}, with the same color code. The abscissa correspond to the voltage applied in the forcing system.}
\label{linear_regime_laminar}
\end{figure*}

This is confirmed in Figure \ref{linear_regime_laminar}, which shows, in linear scale, the response as a function of forcing amplitude at selected frequencies spanning the forcing bandwidths. The filled circles correspond to response amplitudes taken from the spectra of Figure \ref{jet_resp_laminar}, with the same color code. In the band $0.3\leqslant St\leqslant 0.45$ a clear linear regime is present and comparison with Figure \ref{jet_resp_laminar} shows that it is associated with the high values of $\gamma_{dz}$ obtained. This trends continues in the band $0.3\leqslant St\leqslant 0.65$, but saturation occurs at lower forcing amplitudes. For the largest forcing band, no clear linear regime can be identified. The behaviour was the same for other $St$ not shown in the figure. Again, we emphasise the link between the onset of nonlinearity and the drop in coherence (compare, for instance, the saturation of the filled circles with the coherence plots in Figure \ref{jet_resp_laminar}, which are in the same color code).

\begin{figure*}[!ht]
\centering
\includegraphics[trim=4.5cm 0cm 5cm 0cm, clip=true,width=1\linewidth]{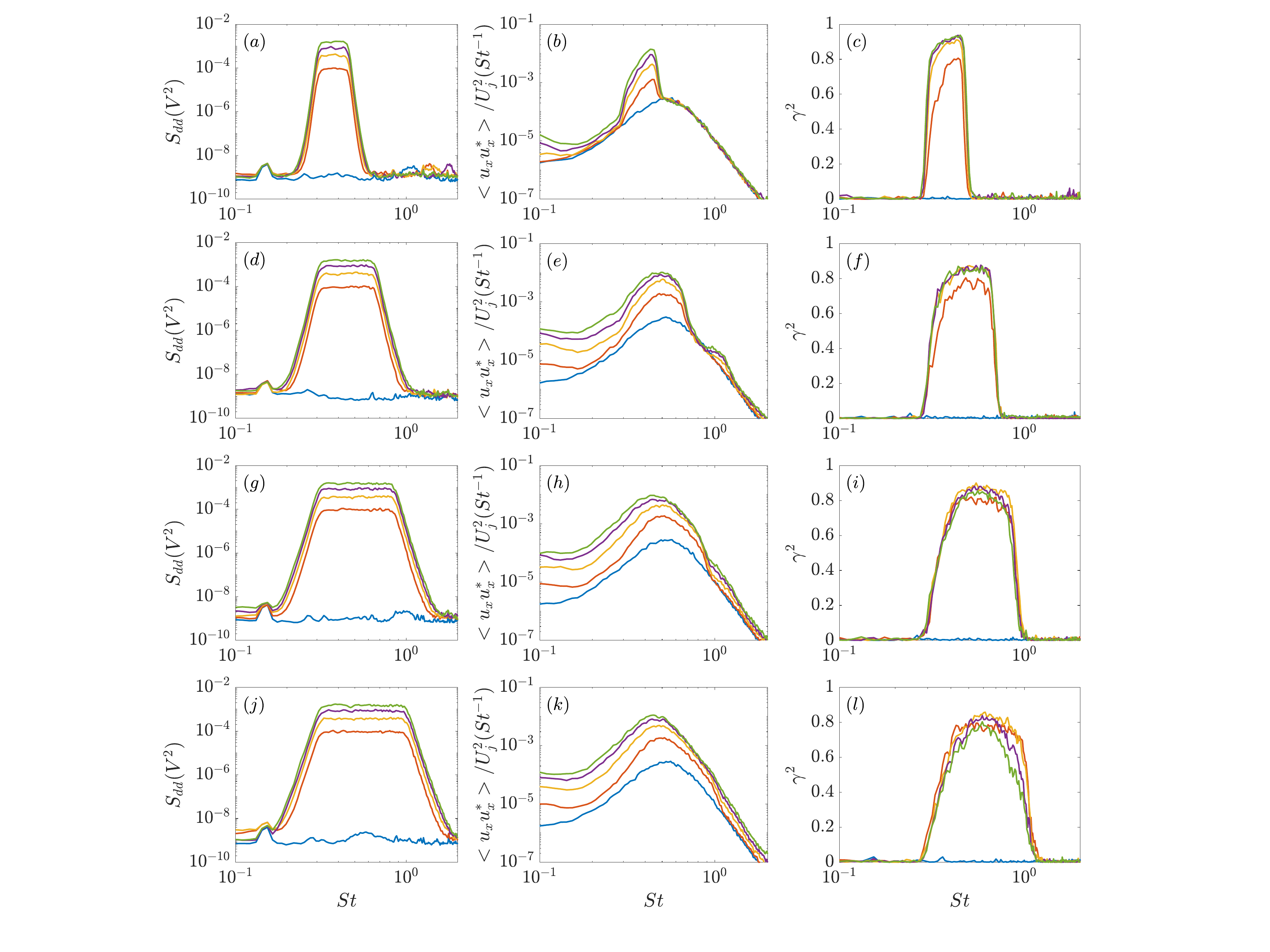}
\caption{Response of the turbulent jet to stochastic forcing in increasing bandwidths. Legend is the same as in Figure \ref{jet_resp_laminar}.}
\label{jet_resp_turbulent}
\end{figure*}

\begin{figure*}[!ht]
\centering
\includegraphics[trim=4.5cm 0cm 5cm 0cm, clip=true,width=1\linewidth]{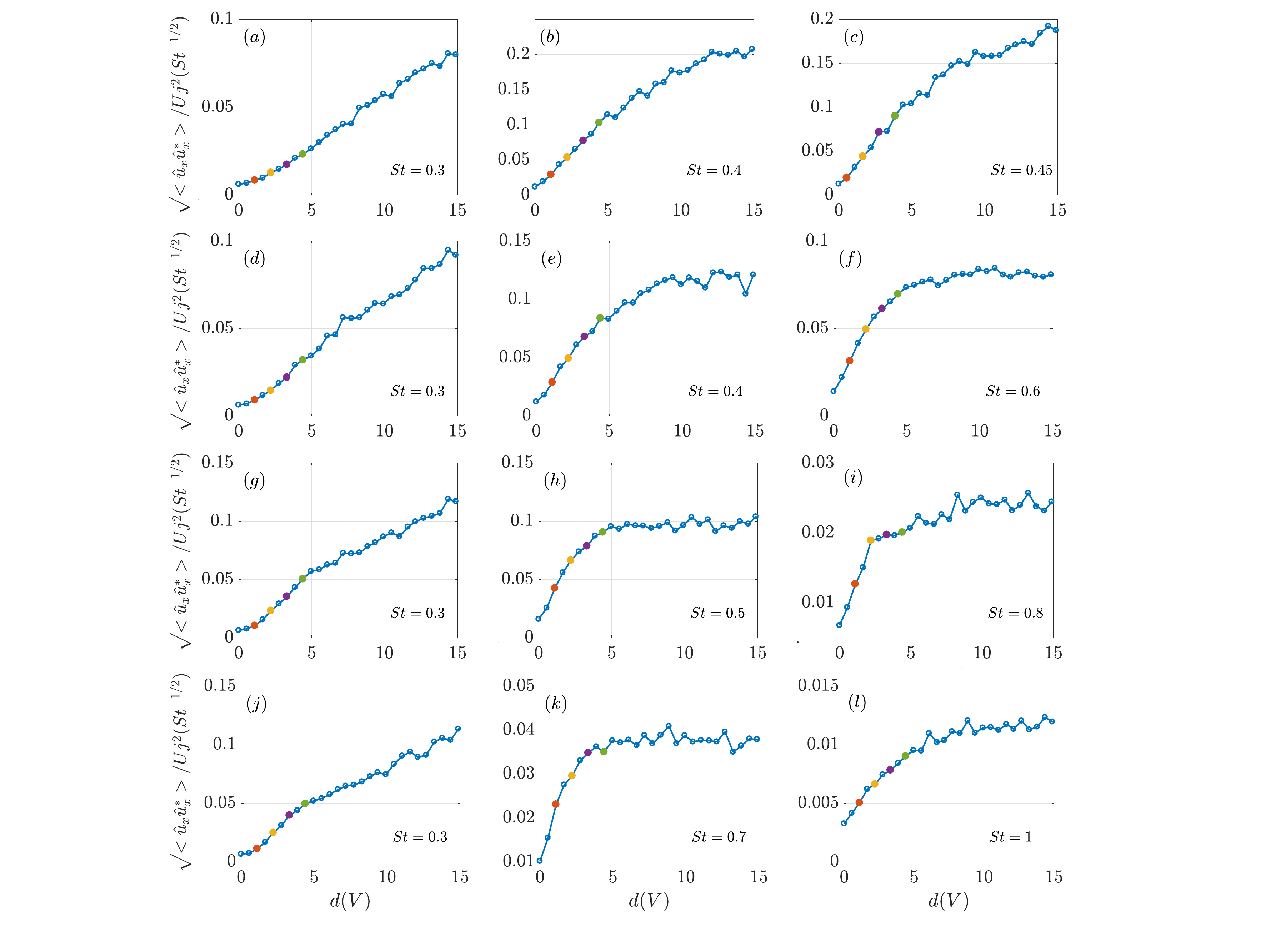}
\caption{Response of the turbulent jet as a function of forcing amplitude for selected frequencies spanning the different forcing bandwidths. Legend is the same as in Figure \ref{linear_regime_laminar}.}
\label{linear_regime_turbulent}
\end{figure*}

Figures \ref{jet_resp_turbulent} and \ref{linear_regime_turbulent} show the same set of plots for the turbulent jet. The forcing amplitudes
and bandwidths are the same used for the initially-laminar jet, allowing a direct comparison between the two cases. Similar to the initially-laminar case, a linear response is obtained for jets forced at the two narrower forcing bandwidths and coherences are as high as 0.9. However, for the two larger bandwidths, there is a clear difference between the initially-turbulent and initially-laminar cases. Saturation occurs at lower amplitudes for the laminar case, as can be seen by comparing figure \ref{linear_regime_laminar} and \ref{linear_regime_turbulent}. In Figure \ref{jet_resp_laminar} we note a superposition of the forced jet spectra in most of the forcing band, which does note happen in the turbulent jet. Furthermore, a consequence of non linearity in the initially-laminar case is the loss of coherence; this phenomenon is much less dramatic in the turbulent case. Indeed, in the bandwidth $0.3 \leqslant St \leqslant 1$, coherence values of $\gamma_{dz}^{2}\approx 0.8$ are observed, as opposed to only $\gamma_{dz}^{2}\approx 0.4$ in the initially-laminar case.

These trends show a clear difference between the responses of initially-laminar and initially-turbulent jets. In the initially-laminar jet, transition takes place close to the nozzle exit, as seen in the rms levels of Figure \ref{jet_val} in Appendix \ref{sec:appendix}. This underpins coherence loss even in the absence of forced wavepackets. If the disturbances are harmonic or narrow-band, then they interact mainly with broadband turbulence; but because the latter has far less energy than the former, the interactions are not sufficiently strong to provoke a desynchronisation of the disturbances, and $\gamma_{dz}$ remains high. If, on the other hand, the forcing bandwidth is large, the number of non-linear interactions between high-amplitude disturbances at different frequencies increases, resulting in phase blur, coherence loss and saturation.

On the other hand, when the jet is initially turbulent the results show that it is easier to maintain a linear relationship between forcing and response. The underlying reason for such contrast between the two cases lies in the lower growth rates of the disturbances on the turbulent shear-layer at higher Strouhal numbers. As seen in Figure \ref{stability_plots}, in the initial jet region and at $St>0.65$ the growth rate of the axisymmetric KH mode is significantly higher in the initially-laminar jet; therefore forcing it at those Strouhal numbers makes nonlinear interactions more prominent.


\subsection{Linearity of actuator-jet response}

Following the trends of Figures \ref{jet_resp_laminar}-\ref{linear_regime_turbulent}, forcing amplitudes were selected for each forcing bandwidth, so as to maintain the jets, as far as is possible, in the linear regime. These amplitudes are then used to produce a baseline forced jet for computation of $H_{dz}$ and to perform the real-time, reactive control experiment. 

\begin{figure}[!ht]
\centering
\includegraphics[trim=2cm 8cm 2cm 6cm, clip=true,width=0.9\linewidth]{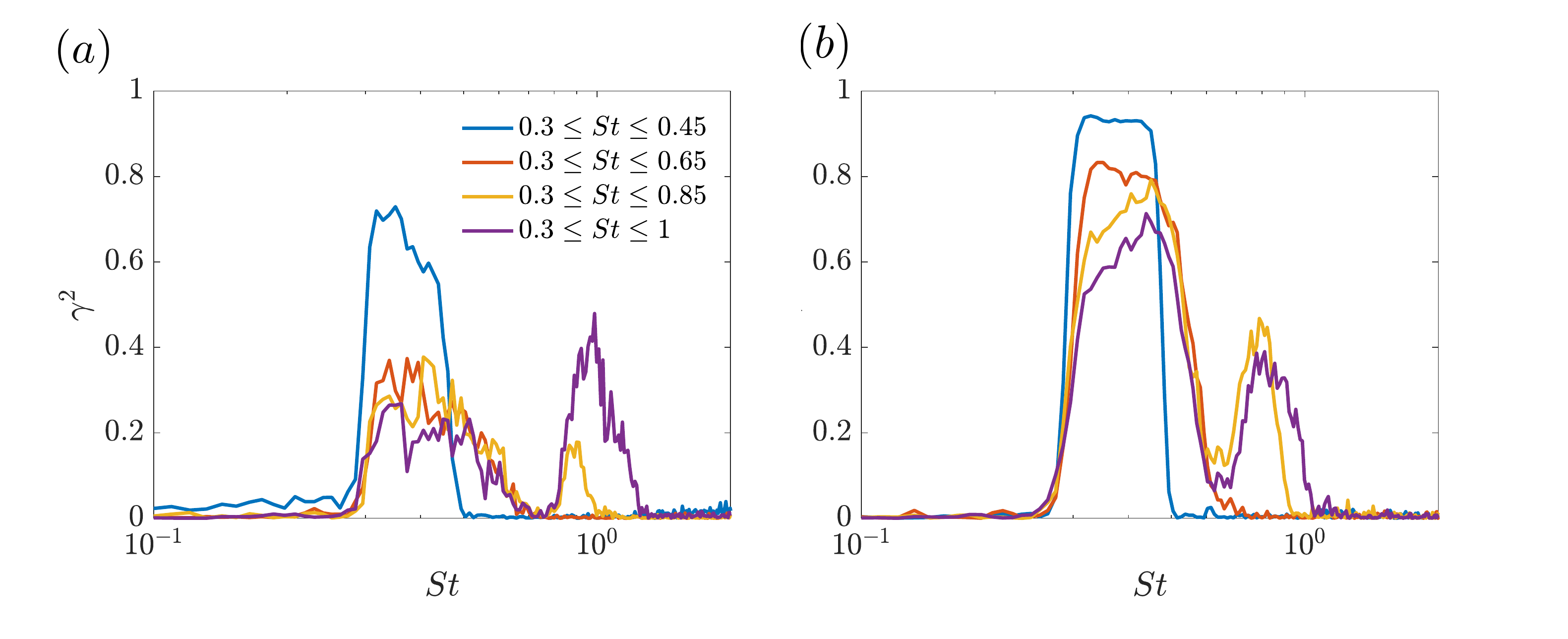}
\caption{Actuator/objective coherence, $\gamma_{uz}$, measured for actuation on the initially-laminar (a) and turbulent (b) jets at different bandwidths.}
\label{coherence_uz_all}
\end{figure}

The actuation amplitudes are then set so that they produce a response at the objective position with amplitudes comparable to those generated by the forcing, for each bandwidth. Figure \ref{coherence_uz_all} shows the actuator/objective coherences, $\gamma_{uz}$, for the initially-laminar and turbulent jets with the actuation amplitudes selected. It can be seen that large actuation bandwidths yield coherence loss. In the laminar case this loss is abrupt between the first and subsequent actuation bands, whereas in the turbulent case the loss is less dramatic, with a trend of systematic decrease in coherence with increase in actuation bandwidth and a sharp decay for $St>0.5$. These issues may be related to either a nonlinear response of the actuators or a nonlinear response of the flow itself. Concerning the latter hypothesis, it should be noted that the radial and streamwise position of the actuators ($x/D=1.5$, $r/D=0.8$) does not correspond to a region of high receptivity in the flow. Furthermore, at this streamwise location, for a range of relevant frequencies, Kelvin-Helmholtz wavepackets are (or are on the verge of becoming) spatially convectively stable, as seen in Figure \ref{stability_plots}. Therefore, in order to produce an actuation signal with high enough amplitudes so as to eliminate the disturbances introduced upstream, one is obliged to increase the amplitude past the linear zone of both actuator and jet, triggering nonlinear effects. This could explain, for instance, the second hump of coherence can be seen in the range $0.7 \leqslant St \leqslant 1$.

The present placement of actuators was chosen in order to accommodate the sensor array in the reactive control experiment described in \citet{Maiaetal2020_control, MaiaetalPRF2021}. Optimising the actuators, as well as the relative distances of inputs, objective and actuators is something to be considered for future work. Here we focus on assessing the effect of nonlinear behaviour (related to both forcing and actuation), on control performance.

%

\subsection{Reactive wave-cancellation}

Having determined forcing and actuation amplitudes for each frequency band considered, the transfer functions $H_{dz}$ and $H_{uz}$ and control kernel, $K$, can be computed using equations \ref{hdz}, \ref{huz} and \ref{K_ctrl_d}.

As discussed by \citet{SasakiTCFD2018_2}, the advantage of wave-cancellation in comparison to Linear Quadratic Gaussian (LQG) control is that the former provides a clear physical interpretation of the control mechanism induced by actuation: it should correspond to a destructive interference pattern between the wavepackets forced at the nozzle exit and the wavepackets generated by the actuators.

\begin{figure}[!ht]
\centering
\includegraphics[trim=1cm 1cm 2cm 3cm, clip=true,width=0.8\linewidth]{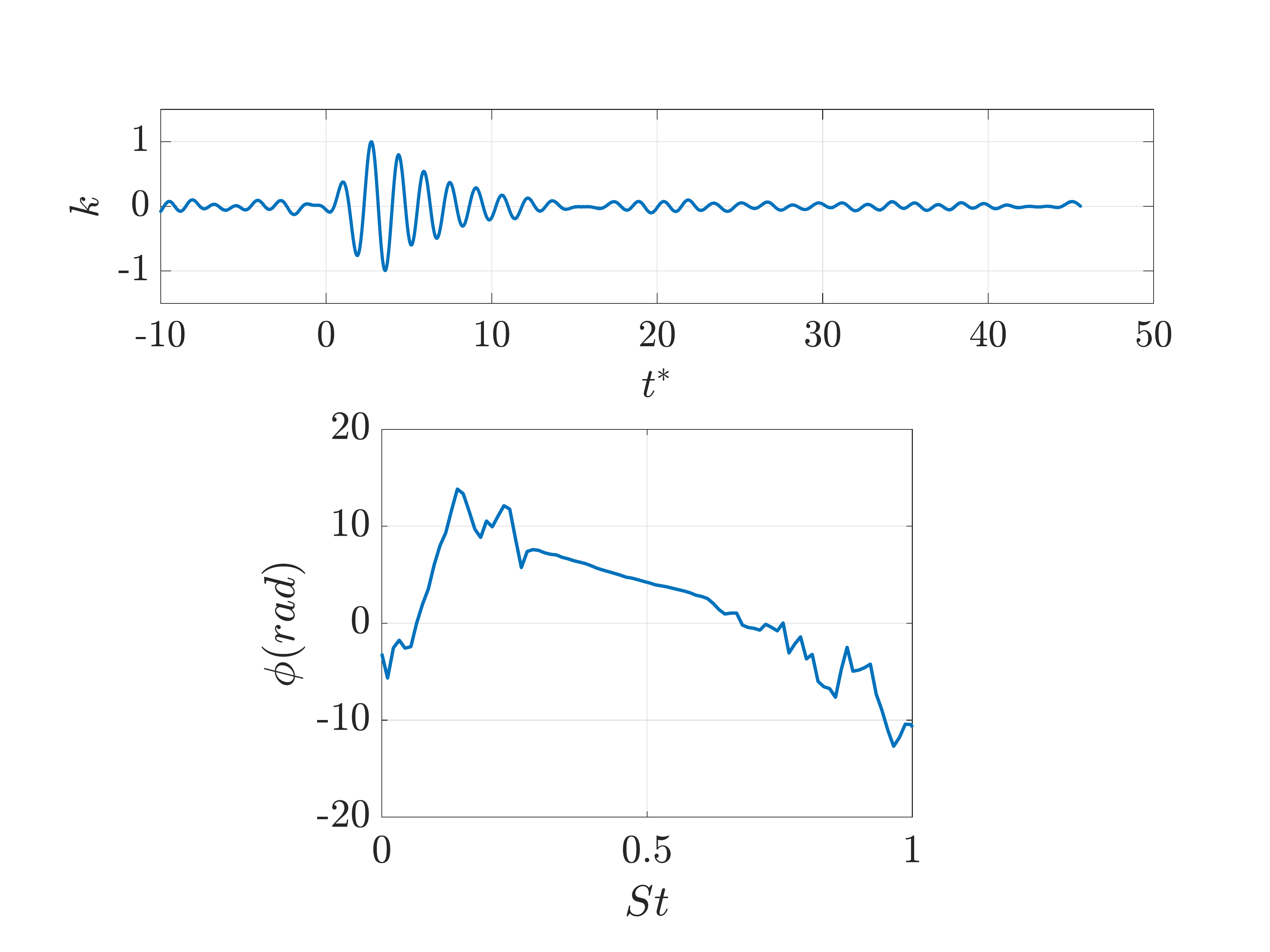}
\caption{Typical control kernel in time domain and its phase as a function of $St$. $t^{*}=tU_{j}/D$ is a non-dimensional time. The kernel was computed for the turbulent jet forced and actuated in the band $0.3 \leqslant St \leqslant 0.65$.}
\label{kernels}
\end{figure}

Figure \ref{kernels} shows an example of a typical kernel, computed for the turbulent jet forced and actuated in the band $0.3\leqslant St \leqslant 0.65$. Only the causal parts ($t^{*}>0$) are used in the convolution. The phases of the kernels as a function of Strouhal number are also displayed. The kernels are shaped like wavepackets characterised by growing and decaying envelopes. In the Strouhal number range where the gains are defined, the phases have linear relationships with $St$. Kernels computed for other forcing and actuation bands in turbulent and initially-laminar jets (not shown) display similar behaviour.

These characteristics confirm that control carried out in this study is underpinned by wave cancellation of the forced wavepackets. 

We assess control performance by comparing PSDs of streamwise velocity fluctuations measured at the objective position of controlled and uncontrolled jets. The baseline for comparison is the jet forced with the amplitudes selected following the procedure described in section \S \ref{sec:seeking}, and the controlled jet both forcing and actuation were active. In order to confirm that real-time control authority is possible, reduction and amplification kernels were computed. The reduction kernel, denoted as $K_{r}$ was computed through equation \ref{K_ctrl_d}; and the amplification kernel, $K_{a}$, was obtained by simply applying a $\pi$ phase shift to $K_{r}$. 

\begin{figure*}[!ht]
\centering
\includegraphics[trim=2cm 2cm 5cm 2cm, clip=true,width=0.8\linewidth]{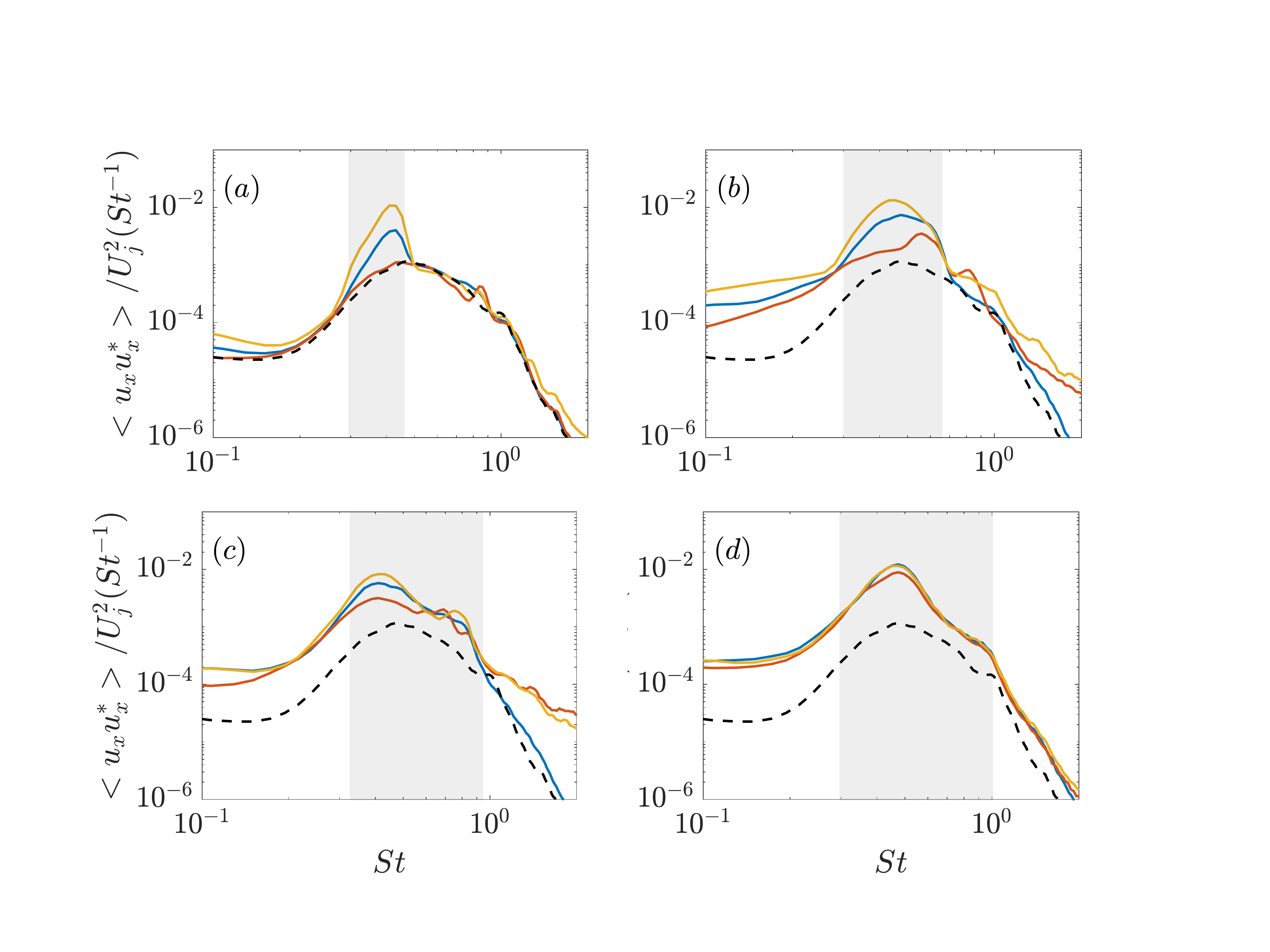}
\caption{Control results for the initially-laminar jet: streamwise velocity PSDs measured at the objective position of controlled and uncontrolled jets. \ref{hwplot1}: Baseline case (forced jet); \ref{hwplot2}: controlled jet with reduction-aimed kernel, $K_{r}$; \ref{hwplot3}: controlled jet with amplification-aimed kernel, $K_{a}$; \ref{hwplot4}: unforced jet. Forcing bandwidths, represented by the gray-shaded areas, are:(a) $0.3 \leqslant St \leqslant 0.45$; (b) $0.3 \leqslant St \leqslant 0.65$; (c) $0.3 \leqslant St \leqslant 0.85$; (d) $0.3 \leqslant St \leqslant 1$.}
\label{control_res_laminar}
\end{figure*}

\begin{figure*}[!ht]
\centering
\includegraphics[trim=2cm 2cm 5cm 2cm, clip=true,width=0.8\linewidth]{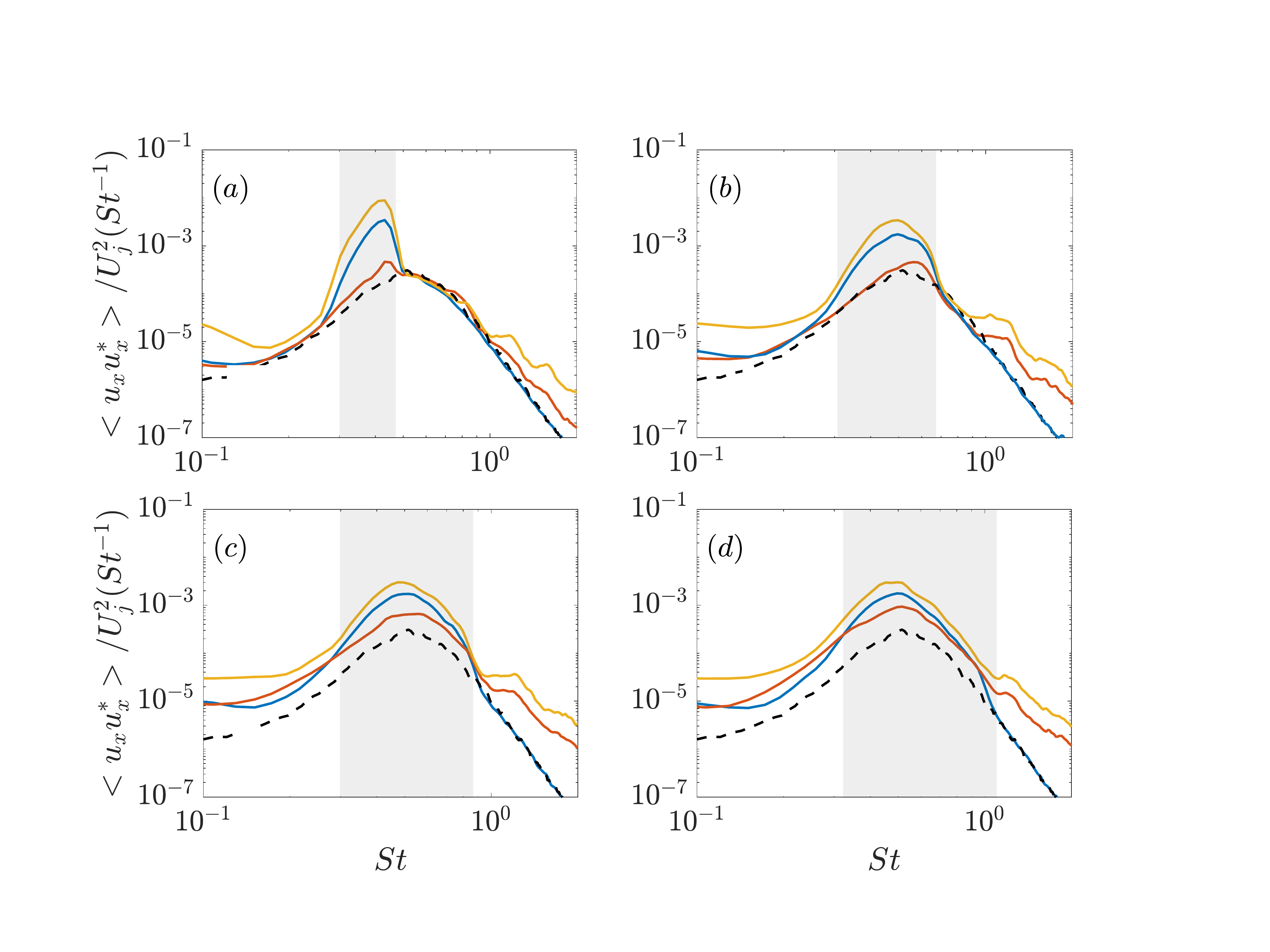}
\caption{Control results for the turbulent jet. Legend is the same as in Figure \ref{control_res_laminar}.}
\label{control_res_turbulent}
\end{figure*}

Figures \ref{control_res_laminar} and \ref{control_res_turbulent} show results of the reactive wave-cancellation experiment in the initially-laminar and turbulent jets, respectively. Data from the unforced jets is also shown for comparison. Concerning the initially-laminar jet, control performed in the narrowest frequency band is successful, in both reducing and amplifying the disturbances, showing that real-time control authority was achieved. The excited wavepackets are virtually eliminated, and the spectrum of the controlled jet resembles closely that of the unforced jet. As forcing bandwith is increased,  control performance is systematically degraded. For the bandwidth $0.3 \leqslant St \leqslant 1$, no clear effect can be distinguished between the baseline and the $K_{r}$ and $K_{a}$ cases.

There is a direct link between the nonlinear effects discussed previously and control performance: for the narrowest band, $\gamma_{dz}$ and $\gamma_{uz}$ are high because nonlinear effects are weak, leading to an effective control. As the frequency band widens, nonlinear effects become more prominent and coherence drops.

In the turbulent jet, for the two narrowest frequency bands of forcing, wave-cancellation is even more effective: reductions of one order of magnitude can be seen, and the spectra of the controlled jets are similar to that of the unforced jet. For the two larger bandwidths performance decreases and amplitudes could not be reduced to the unforced jet levels. There is no significant drop in $\gamma_{dz}$ with increasing bandwidth, as can be seen in Figure \ref{jet_resp_turbulent}, which means that  wavepacket estimation remains accurate. The reduction in control efficiency is therefore underpinned by the decaying values of $\gamma_{uz}$. Despite this issue, control has a very clear effect up until the widest frequency band of forcing.

Comparison of Figures \ref{control_res_laminar} and \ref{control_res_turbulent} reveals that wave-cancellation results, for most of the frequency bands tested, are better in the turbulent case, consistent with the open-loop results which showed that the linear regime is more easily maintained in the initially turbulent jet, therefore providing a more favourable scenario for wave-cancellation. 

\scalebox{0}{%
\begin{tikzpicture}
    \begin{axis}[hide axis]
        \addplot [
        color=color1,
        solid,
        line width=1.2pt,
        forget plot
        ]
        (0,0);\label{hwplot1}
        \addplot [
        color=color2,
        solid,
        line width=1.2pt,
        forget plot
        ]
        (0,0);\label{hwplot2}
        \addplot [
        color=color3,
        solid,
        line width=1.2pt,
        forget plot
        ]
        (0,0);\label{hwplot3}
        \addplot [
        color=black,
        dashed,
        line width=1.2pt,
        forget plot
        ]
        (0,0);\label{hwplot4}
    \end{axis}
\end{tikzpicture}%
}%

\section{Conclusions}\label{sec:conclusions}

An experiment study on the response of stochastically-forced, initially-laminar and initially-turbulent jets has been performed with a view to assessing the performance of wave-cancellation-based reactive control. The study is an extension of recent work that have reported a successful implementation of real-time reactive control in forced turbulent jets through distructive interference \cite{Maiaetal2020_control,MaiaetalPRF2021}.

The jet is forced by disturbances introduced at the nozzle lip so as to raise wavepackets levels above that of background turbulence, thus simplifying the task of identifying and controlling the disturbances. The control strategy is a simplification of the inverse feedforward scheme of \cite{SasakiTCFD2018_2, SasakiTCFD2018_1}, insofar as we use the prescribed disturbances as input for the control law instead of flow measurements, a strategy also applied in \citet{Maiaetal2020_control,MaiaetalPRF2021}. Control law design relies on input-output transfer functions, which are computed empirically by measuring the response of the jet to forcing and actuation separately.

Our analysis of the open-loop response of the jets to stochastic forcing builds on a significant body of work that has been dedicated to harmonically forced jets. Given the importance of the linear regime for wave-cancellation-based control, we characterise the onset of nonlinear response in the jet for different frequency bands of forcing and use this to understand control performance. Special attention is paid to the differences between forcing and control in jets with laminar and turbulent boundary layers.

The results show that, when forcing is narrow-band, there is a clear linear response regime, both for initially-laminar and initially-turbulent jets. However, as forcing bandwidth is increased, nonlinear effects are activated. These effects were found to be more prominent in the initially-laminar jet, in which disturbances undergo higher growth rates in the initial region and transition to turbulence. Similar trends were observed in the jet response to actuation. The results also show that departure from linearity is systematically followed by a loss of coherence between disturbance/actuator and objective.

These issues directly impact control performance in the initially-laminar jet. While at narrow frequency bands of forcing and actuation wave cancellation is very effective, as bandwidths are increased control performance is significantly degraded. The turbulent jet, on the other hand, offered better conditions for application of wave cancellation, on account of lower growth rates of disturbances at a broad range of Strouhal numbers. It was possible to increase the bandwidth of the forced wavepackets without triggering significant nonlinear behaviour, and the control produced a very clear effect up until the largest frequency band tested, $0.3 \leqslant St \leqslant 1$. 

\section{Supplementary material}
Images, videos and more details about the PIV setup are available at \href{url}{https://doi.org/10.6084/m9.figshare.13048586}.

\begin{acknowledgements}
I. A. M. acknowledges support from the Science Without Borders program through the CNPq grant number 200676/2015-6. P. J and A. V. G. C. acknowledge support from the CAPES Science Without Borders project no. A073/2013. The authors wish to thank Redouane Kari and Damien Eysseric for their invaluable work during the experimental campaign.
\end{acknowledgements}

\appendix

\section{Turbulent boundary layer}
\label{sec:turb_bl}

\begin{figure}
\centering
\includegraphics[trim=0cm 0cm 0cm 0cm, clip=true,width=0.5\linewidth]{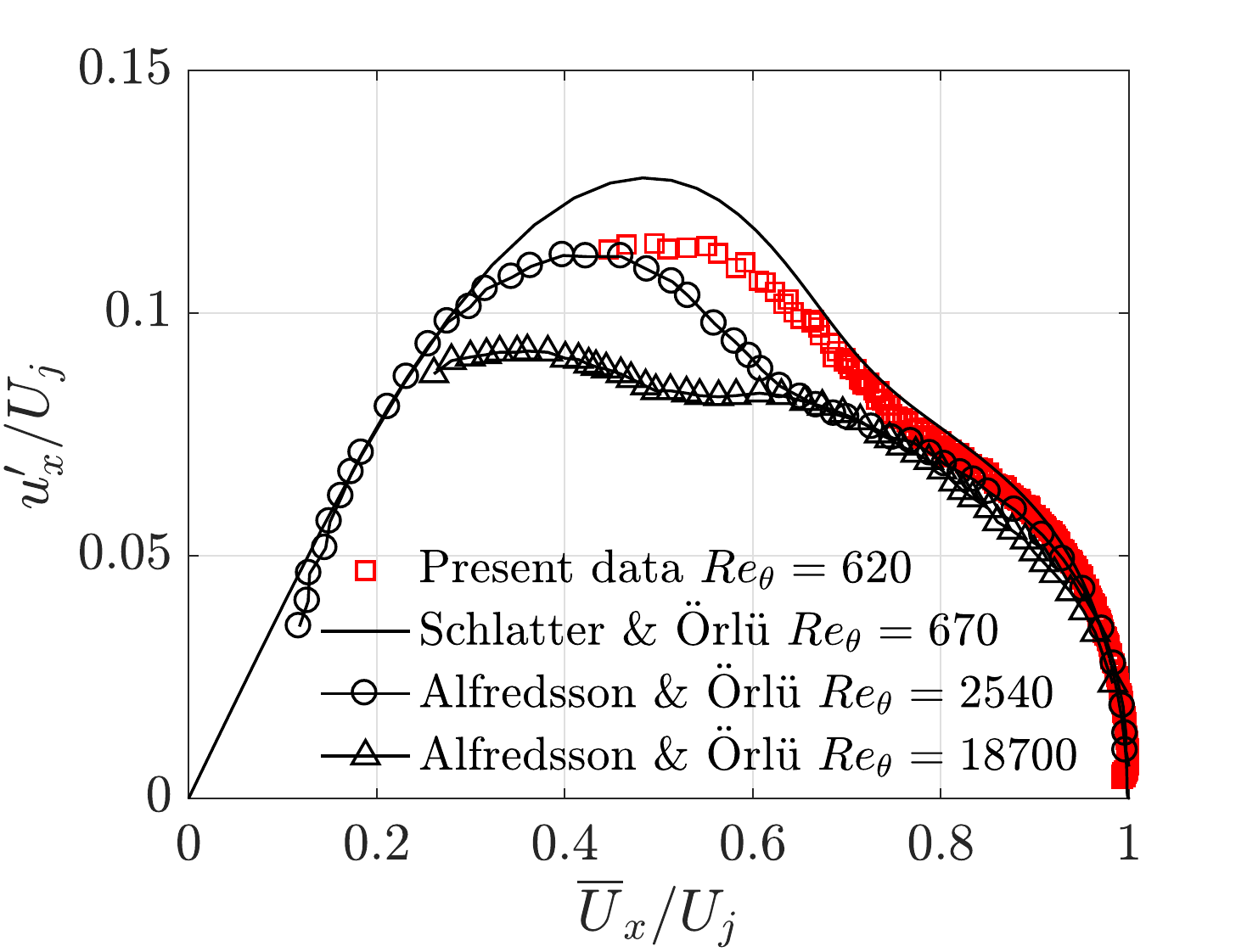}
\caption{Boundary layer diagnostic plot. Our data is compared to experiments from \citet{Orlu2010_2} at higher $Re_{\delta_{2}}$ and direct numerical simulation (DNS) data from \citet{SchlatterOrlu} for a similar $Re_{\delta_{2}}$.}
\label{bl_diagnostic_plot}
\end{figure}

In this section we present a comparison between the tripped boundary layer profiles and data from the literature for turbulent boundary layers. We make the comparison in base on the \textit{diagnostic plot} proposed by \citet{Alfredsson2010}, which is independent of wall position. Figure \ref{bl_diagnostic_plot} shows a comparison of diagnostic plots between our data and experimental results by \cite{Orlu2010_2} for higher $Re_{\delta_{2}}$. Our data follows two trends expected of turbulent boundary layers: decreasing peak rms, $u'_{x}/U_{j}$ with increasing $Re_{\delta_{2}}$ and a shift this peak towards lower values of $\bar{U}_{x}/U_{j}$ \citep{Alfredsson2010}. This shows that the
tripping successfully leads to a fully-turbulent condition at the nozzle exit.



\section{Aerodynamic measurements}
\label{sec:appendix}

\begin{figure*}
\centering
\includegraphics[trim=0cm 0cm 0cm 0cm, clip=true,width=0.9\linewidth]{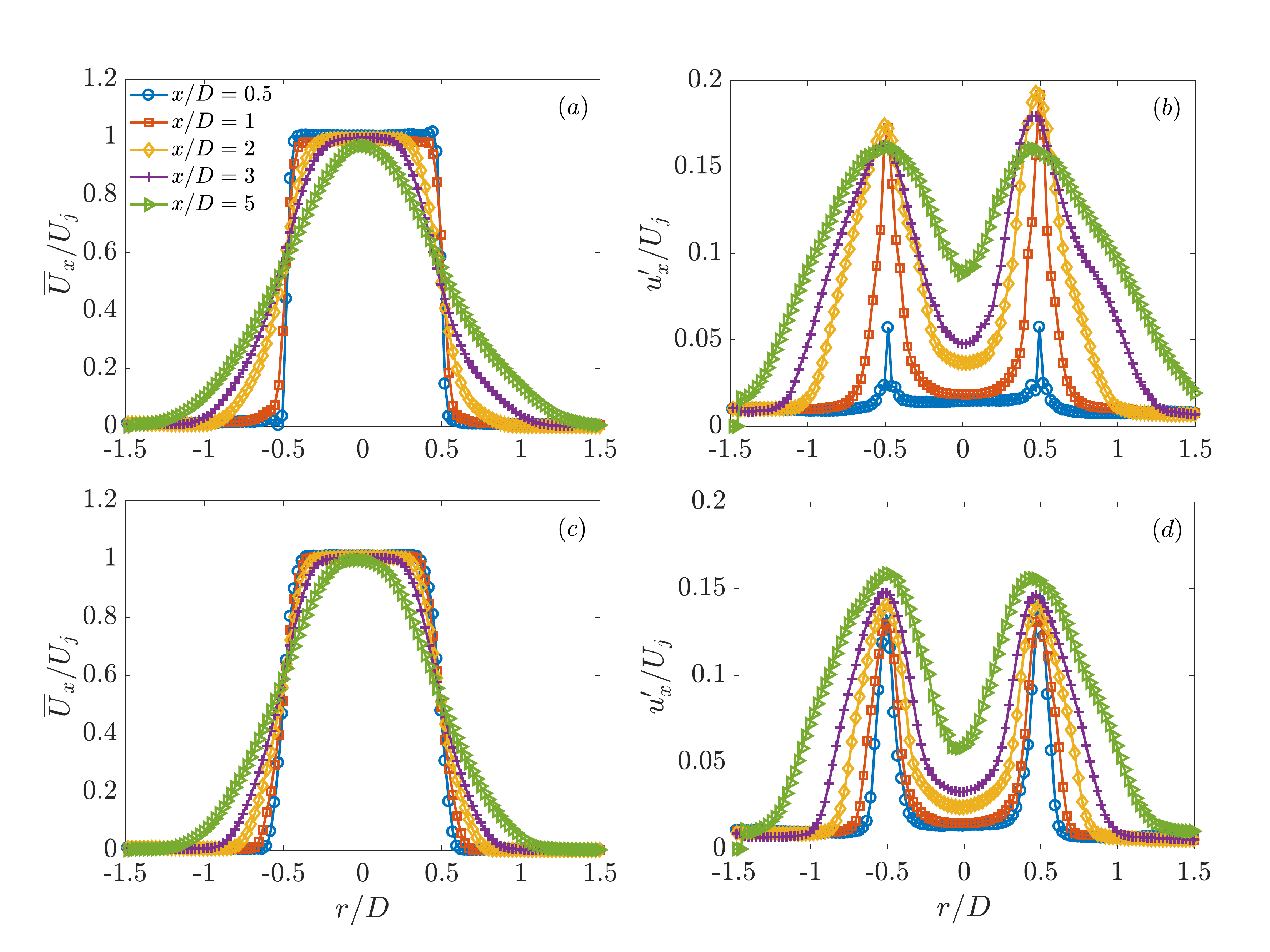}
\caption{Mean and rms radial velocity profiles at different streamwise positions. (a) and (b): initially-laminar jet; (c) and (d): turbulent jet.}
\label{jet_val}
\end{figure*}

An aerodynamic characterisation was carried out in order to compare jets with tripped and untripped boundary layers. Figure \ref{jet_val} shows mean and rms radial profiles of streamwise velocity at different streamwise positions for the initially laminar and turbulent jets. The streamwise evolution from a top-hat to a bell-shaped profile can be seen. In the initial region of the laminar jet, we note that the turbulence intensity, $u_{x}'/U_{j}$, goes from very low values at $x/D=0.5$ to a peak at $x/D=2$ and a subsequent decrease. This behaviour can be associated with transition to turbulence \citep{Bresetal2018}, and
it can be noted that it happens in a slightly asymmetric fashion. In the fully turbulent jet, on the other hand, the profiles are symmetric
for all streamwise positions measured, with a peak turbulence intensity of approximately 0.16.

\section{Linearised Navier-Stokes equations}
\label{sec:linearised_NS}

The linearised Navier-Stokes equations for axisymmetric ($m=0$) disturbances are given by:

\begin{equation}
\begin{split}
&-i\omega \hat{u}_{x} + i\alpha \bar{U}_{x}\hat{u}_{x} +\hat{u}_{r}\bar{U}_{x}'=-\frac{\gamma-1}{\gamma}i\alpha \hat{T} -\frac{\gamma-1}{\gamma}\frac{\bar{T}}{\bar{\rho}}i\alpha \hat{\rho} +\frac{1}{\bar{\rho}Re}\left[\bar{\mu}\left(-\alpha^2\hat{u}_{x}+ \frac{1}{r}\frac{\partial}{\partial r}\left(r\frac{\partial \hat{u}_{x}}{\partial r}\right)\right) \biggr. \right. \\ & +\frac{\bar{\mu}}{3}\biggl( -\alpha^2\hat{u}_{x} + \left. \left.i\alpha\frac{\partial \hat{u}_{r}}{\partial r}+\frac{i\alpha}{r} \hat{u}_{r}\right) \right.+ \frac{\partial \bar{\mu}}{\partial r}\left(i\alpha \hat{u}_{r}+\frac{\partial \hat{u}_{x}}{\partial r}\right) +\frac{\mu}{r}\frac{\partial}{\partial r}\left(r \frac{\partial \bar{U}_{x}}{\partial r}\right) \left. \frac{\partial \mu}{\partial r}\frac{\partial \bar{U}_{x}}{\partial r}\right], 
\end{split}
\label{x_momentum}
\end{equation}

\begin{equation}
\begin{split}
&-i\omega \hat{u}_{r} + i\alpha \bar{U}_{x}\hat{u}_{r}=-\frac{\gamma-1}{\gamma}\frac{\partial \hat{V}}{\partial r}-\frac{\gamma-1}{\gamma}\frac{\bar{T}}{\bar{\rho}}\frac{\partial \hat{\rho}}{\partial r} -\frac{\gamma-1}{\gamma}\frac{\hat{\rho}}{\bar{\rho}}\frac{\partial \bar{T}}{\partial r}-\frac{\gamma-1}{\gamma}\frac{\hat{T}}{\bar{\rho}}\frac{\partial \bar{\rho}}{\partial r} + \frac{1}{\bar{\rho}Re}\biggl[\bar{\mu}\biggl(-\alpha^2\hat{u}_{r} \frac{1}{r}\frac{\partial}{\partial r}\left(r\frac{\partial \hat{u}_{r}}{\partial r}\right)\biggr)\biggr. \\ & \left. -\frac{\bar{\mu}\hat{u}_{r}}{r^2} + i\alpha\mu\bar{U}_{x}'+\frac{\bar{\mu}}{3}\left(i\alpha \frac{\partial \hat{u}_{x}}{\partial r} +\frac{\partial^2\hat{u}_{r}}{\partial r^2}+\frac{1}{r}\frac{\partial \hat{u}_{r}}{\partial r}-\frac{\hat{u}_{r}}{r^2} \right)+\frac{\partial \bar{\mu}}{\partial r}\left(2\frac{\partial \hat{u}_{r}}{\partial r}-\frac{2}{3}i\alpha \hat{u}_{x} -\frac{2}{3}\frac{\partial \hat{u}_{r}}{\partial r}-\frac{2}{3}\frac{\hat{u}_{r}}{r}\right)\right],
\end{split}
\label{r_momentum}
\end{equation}

\begin{equation}
\begin{split}
&-i\omega\hat{T}+\hat{u}_{r}\frac{\partial \bar{T}}{\partial r} +i\alpha\bar{U}_{x}\hat{T} +(\gamma-1)\bar{T}\left(i\alpha{u}_{x} +\frac{\partial \hat{u}_{r}}{\partial r}+\frac{\partial \hat{u}_{r}}{\partial r}\right) =\frac{\gamma}{\bar{\rho}RePr}\left[\bar{\mu}\left(-\alpha^2\hat{T}+\frac{\partial^2 \hat{T}}{\partial r^2}+\frac{1}{r}\frac{\partial \hat{T}}{\partial r}\right)  +\frac{\partial \bar{\mu}}{\partial r}\frac{\partial \hat{T}}{\partial r}  \right. \\ &+\frac{\partial \mu}{\partial r}\frac{\partial \bar{T}}{\partial r}  +\left.\mu\left(\frac{\partial^2 \bar{T}}{\partial r^2}+\frac{1}{r}\frac{\partial \bar{T}}{\partial r}\right)\right] + \frac{\gamma}{\bar{\rho}}\left[\frac{\mu}{Re}\left(\frac{\partial \bar{U}_{x}}{\partial r}\right)^2   \right. + \left. \frac{2\bar{\mu}}{Re}\frac{\partial \bar{U}_{x}}{\partial r}\left(i\alpha\hat{u}_{r}+\frac{\partial \hat{u}}{\partial r}\right) \right],
\end{split}
\label{energy}
\end{equation}

\begin{equation}
-i\omega\hat{\rho}+i\alpha\bar{U}_{x}\hat{\rho}+\bar{\rho}i\alpha \hat{u} +\bar{\rho}\frac{\partial \hat{v}}{\partial r}+\hat{v}\frac{\partial \bar{\rho}}{\partial r} +\frac{\bar{\rho}\hat{v}}{r}=0,
\label{continuity}
\end{equation}
for $x$-momentum, $r$-momentum, energy and continuity, respectively. The hats denote Fourier transformed quantities. The mean fields of temperature and density are determined using the Crocco-Busemann relation and the perfect gas law. The viscosity, $\bar{\mu}$ is determined using Sutherland's law, and the bulk viscosity has been assumed to be zero. The coefficients are then rearranged in matrices $\textbf{L}$ and $\textbf{F} $ shown in equation \ref{gen_eigenvalue} to yield an eigenvalue problem.

\bibliographystyle{unsrtnat}
\bibliography{bibfile}

\end{document}